\documentclass[conference]{IEEEtran}
\IEEEoverridecommandlockouts
\usepackage{cite}
\usepackage{amsmath,amssymb,amsfonts}
\usepackage{algorithmic}
\usepackage{graphicx}
\usepackage{textcomp}
\usepackage{xcolor}
\usepackage{booktabs}
\usepackage[hidelinks]{hyperref}
\usepackage{booktabs} 
\usepackage{array}    

\usepackage{arydshln}
\usepackage{graphicx}
\usepackage{subcaption} 
\usepackage{comment}

\def\BibTeX{{\rm B\kern-.05em{\sc i\kern-.025em b}\kern-.08em
    T\kern-.1667em\lower.7ex\hbox{E}\kern-.125emX}}
\begin{document}

\title{One Model, Many Latencies: Universal Speech Enhancement for
Diverse Real-Time Applications
}


\author{
\IEEEauthorblockN{
Szu-Wei Fu\IEEEauthorrefmark{1},
Rong Chao\IEEEauthorrefmark{2},
Xuesong Yang\IEEEauthorrefmark{1},
Sung-Feng Huang\IEEEauthorrefmark{1}, \\
Ante Jukić\IEEEauthorrefmark{1},
Yu Tsao\IEEEauthorrefmark{2},
Yu-Chiang Frank Wang\IEEEauthorrefmark{1}
}

\IEEEauthorblockA{\IEEEauthorrefmark{1}NVIDIA}

\IEEEauthorblockA{\IEEEauthorrefmark{2}Academia Sinica, Taipei, Taiwan}
}

\maketitle

\begin{abstract}
Different real-time speech applications impose distinct latency budgets, often requiring separately trained enhancement models for each scenario. In this paper, we propose a one-for-all, real-time universal speech enhancement model that provides explicit control over both algorithmic and computational latency. Algorithmic latency is flexibly adjusted via configurable look-ahead frames. To avoid learning inefficiency caused by varying padding configurations, we introduce parallel convolutional layers corresponding to different look-ahead settings. Computational latency is controlled through an early-exit mechanism, enabling inference at different network depths. To narrow the performance gap between specialized and flexible models, we propose a two-stage training strategy with a shared-to-multiple decoder transition. Overall, the proposed framework enables a single model to be deployed across diverse latency budgets without retraining separate models. Model weights are available for download at: \url{https://huggingface.co/nvidia/Real-time_RE-USE}
\end{abstract}

\begin{IEEEkeywords}
universal speech enhancement,  real-time processing, latency budget
\end{IEEEkeywords}

\section{Introduction}
\label{introduction}
Recent studies in speech enhancement (SE) have increasingly moved beyond task-specific approaches toward unified models that generalize across heterogeneous domains~\cite{liu2021voicefixer, serra2022universal, li2024masksr, babaev2024finally, zhang2025anyenhance, karita2025miipher, nakata2025sidon}. In this context, universal speech enhancement (USE) seeks to improve intelligibility and perceptual quality under diverse degradation conditions while preserving intrinsic attributes such as speaker identity, emotion, and accent. Although Miipher-2~\cite{karita2025miipher}, and RE-USE \cite{fu2025you}
have demonstrated effectiveness in improving training data quality for other speech generative models (e.g., text-to-speech), their non-causal architectures limit their applicability to real-time scenarios.

\begin{figure}[ht]
  \centering
  \includegraphics[width=\linewidth]{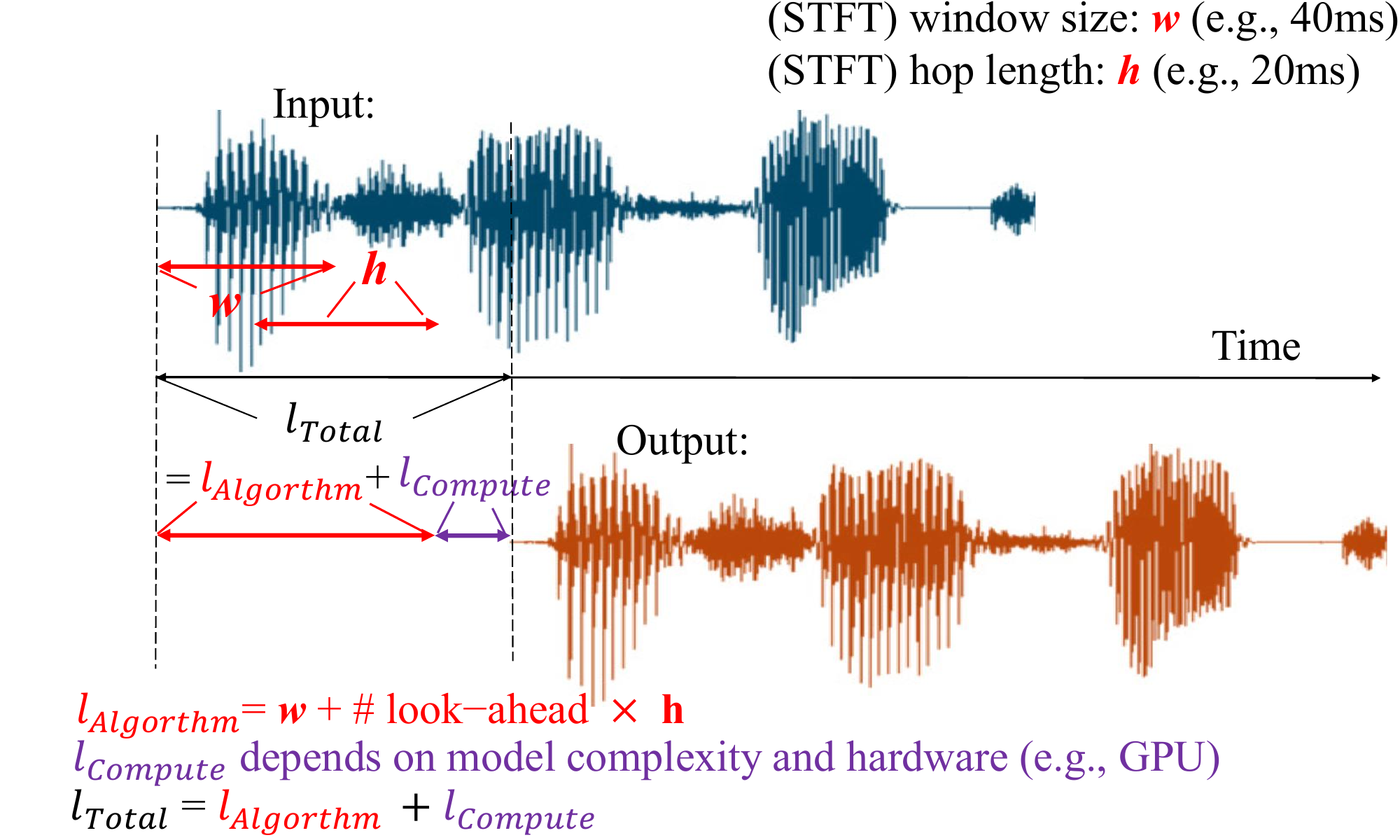}
  \caption{The latency of a speech enhancement system can be categorized into algorithmic latency and computational latency.}
  \label{fig:latency_def}
\end{figure}

The total latency of a real-time speech
enhancement model is the \textbf{sum of algorithmic latency and computational latency}, as illustrated in Figure~\ref{fig:latency_def}. Algorithmic latency refers to the amount of input required by the model to produce the first output unit. For frequency-domain–based USE models, the \textbf{algorithmic latency equals the STFT window size $w$ plus the product of the look-ahead frames and the hop size $h$}. Computational latency denotes the time required by the model to generate an output after receiving the necessary input data, and \textbf{it depends on both model complexity and the computational capability of the deployment hardware}.

Unlike the restoration of pre-recorded speech, which prioritizes output quality over latency constraints, real-time speech enhancement must operate under strict latency budgets. These constraints vary across applications: 
interactive speech applications such as conversational VoIP typically tolerate 50–150 ms~\cite{kuhn2005security}, while streaming ASR systems generally operate within latency budgets of 100–200 ms~\cite{song2023trimtail}. Beyond different latency budget considerations, computational latency is influenced by the computational power of the deployment hardware (see Figure~\ref{fig:latency_def}). \textbf{These factors prevent a single causal model from being suitable across different applications.} Accordingly, this paper aims to develop a training method that allows a single model to accommodate varying deployment conditions, such as latency constraints and hardware specifications.

To enable a single streaming ASR model to support various latency settings, \cite{noroozi2024stateful} proposed a multiple look-ahead training strategy by randomly sampling the chunk size.
An effective strategy for reducing computational latency is early exit~\cite{teerapittayanon2016branchynet}, a training paradigm that enables a network to produce predictions at intermediate layers rather than only at the final layer, allowing the model to seamlessly adapt to different deployment conditions during inference. In the context of speech enhancement, several studies~\cite{li2021learning, chen2021don, kim2022bloom, miccini2023dynamic, feng2025towards, olsen2025knowing} have adopted early-exit mechanisms to build flexible models. Nevertheless, most of them focus primarily on designing the exit criteria. For example, Li et al.~\cite{li2021learning} and Chen et al.~\cite{chen2021don} explored exit strategies based on the similarity between outputs of consecutive layers, using predefined thresholds to determine when to exit. Beyond dynamically adjusting the model depth,~\cite{feng2025towards} introduces FlexAttention to enable flexible control over model width.

Although early-exit can adjust computational latency during inference, algorithmic latency remains fixed. In this paper, we propose a \textbf{one-for-all}, real-time, streamable USE model that not only handles diverse degradation conditions but also provides explicit control over algorithmic latency via flexible look-ahead frames, thereby enabling adaptable total latency and facilitating deployment across a wide range of latency budgets.  

\section{From Offline to Real-Time Speech Enhancement}
\label{Real-Time requirements}

Before presenting our proposed method, this section outlines the constraints that a real-time SE model must satisfy, based on the total latency definition illustrated in Figure~\ref{fig:latency_def}. Some of these constraints are often overlooked in prior studies. Specifically, real-time SE models must meet the following three requirements:

1. \textbf{Causality}: The model architecture should be causal or allow only a limited number of look-ahead frames.

2. \textbf{Latency budget} ($l_{\text{budget}}$): The total latency must not exceed the application-specific latency budget (examples are discussed in Section~\ref{introduction}):
\begin{equation}
\label{eq1}
l_{\text{Total}} \leq l_{\text{budget}}.
\end{equation}

3. \textbf{Real-time processing}: To prevent latency from accumulating over time, the computation time per processing step must be shorter than the corresponding hop size $h$ (i.e., the real-time factor (RTF) must be less than 1):
\begin{equation}
\label{eq2}
l_{\text{Compute}} \leq h.
\end{equation}

As noted by \cite{welker2025real}, some prior studies may report RTF measured under offline processing, where whole utterances are processed at once. This setting exploits time-dimension parallelism and highly optimized large-tensor CUDA kernels, and thus can substantially underestimate the true RTF in streaming inference, obscuring whether a method is practically real-time. 

\section{Proposed Method}

\subsection{Adjustable Algorithmic Latency}

\begin{figure}[t!]
  \centering
  \includegraphics[width=\linewidth]{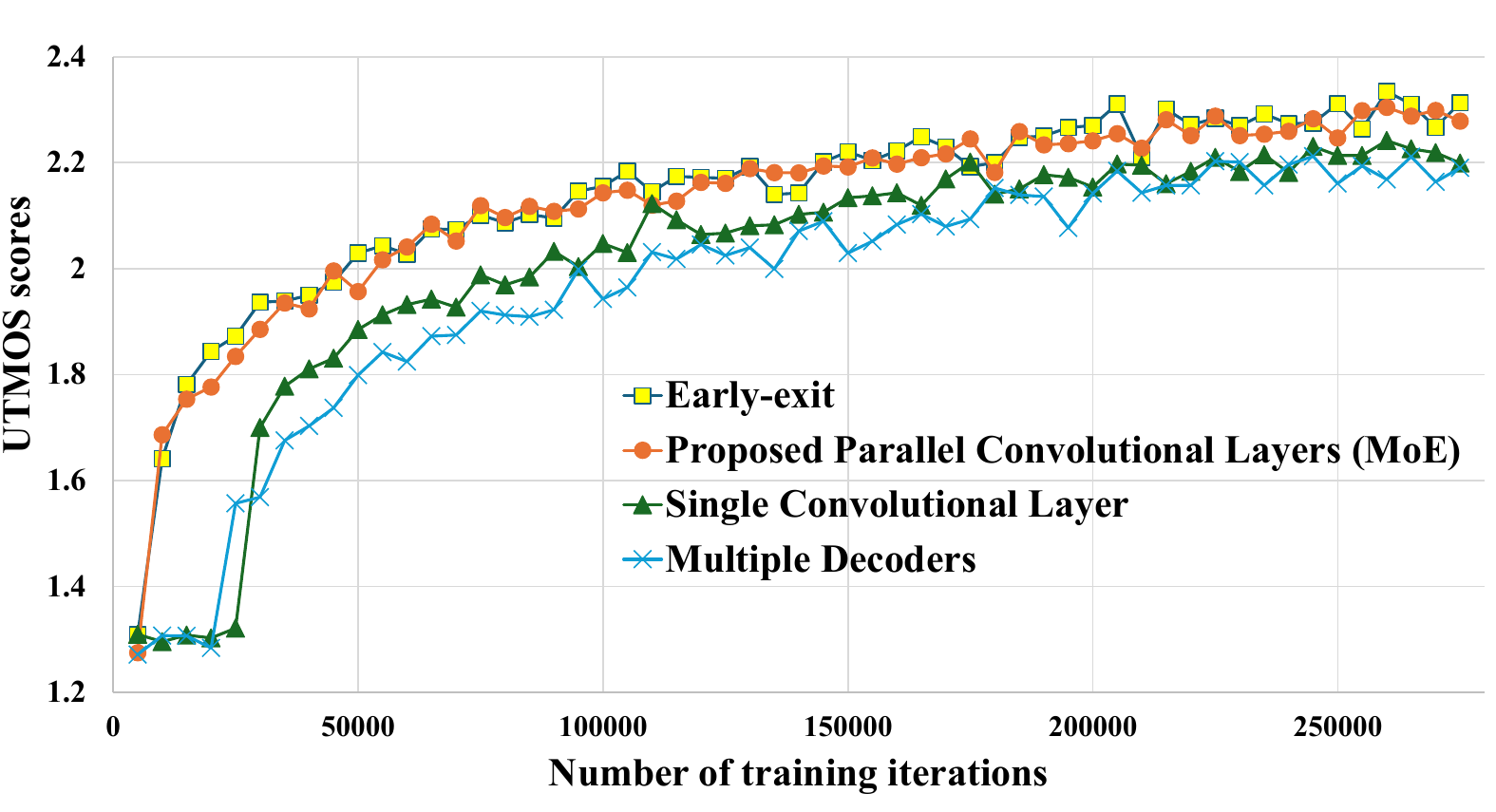}
  \caption{Learning curves of UTMOS scores on the validation set under different model architectures.}
  \label{fig:Learnining_curves}
\end{figure}

To enable adjustable algorithmic latency during inference, it is more practical to control the number of look-ahead frames rather than modifying the STFT window size or hop length. In practice, the number of look-ahead frames can be determined by appropriately setting the left and right padding of the convolutional layers. For example, consider a convolutional layer with a kernel size of 3. To achieve 0, 1, and 2 look-ahead frames, the (left padding, right padding) number can be set to (2, 0), (1, 1), and (0, 2), respectively. However, since convolution is \textbf{translation equivariant} (i.e., if $f$ is a convolution operation and $T$ is a translation (shift) operator: $f(T(x)) = T(f(x))$), using a single convolutional layer with varying padding configurations may disrupt the following sequence modeling and thereby reduce the model’s learning efficiency (see the green learning curve corresponding to the UTMOS score on the validation set in Figure~\ref{fig:Learnining_curves}. Detailed experimental settings are provided in Section~\ref{exp}). 

\begin{figure}[t!]
  \centering
  \includegraphics[width=\linewidth]{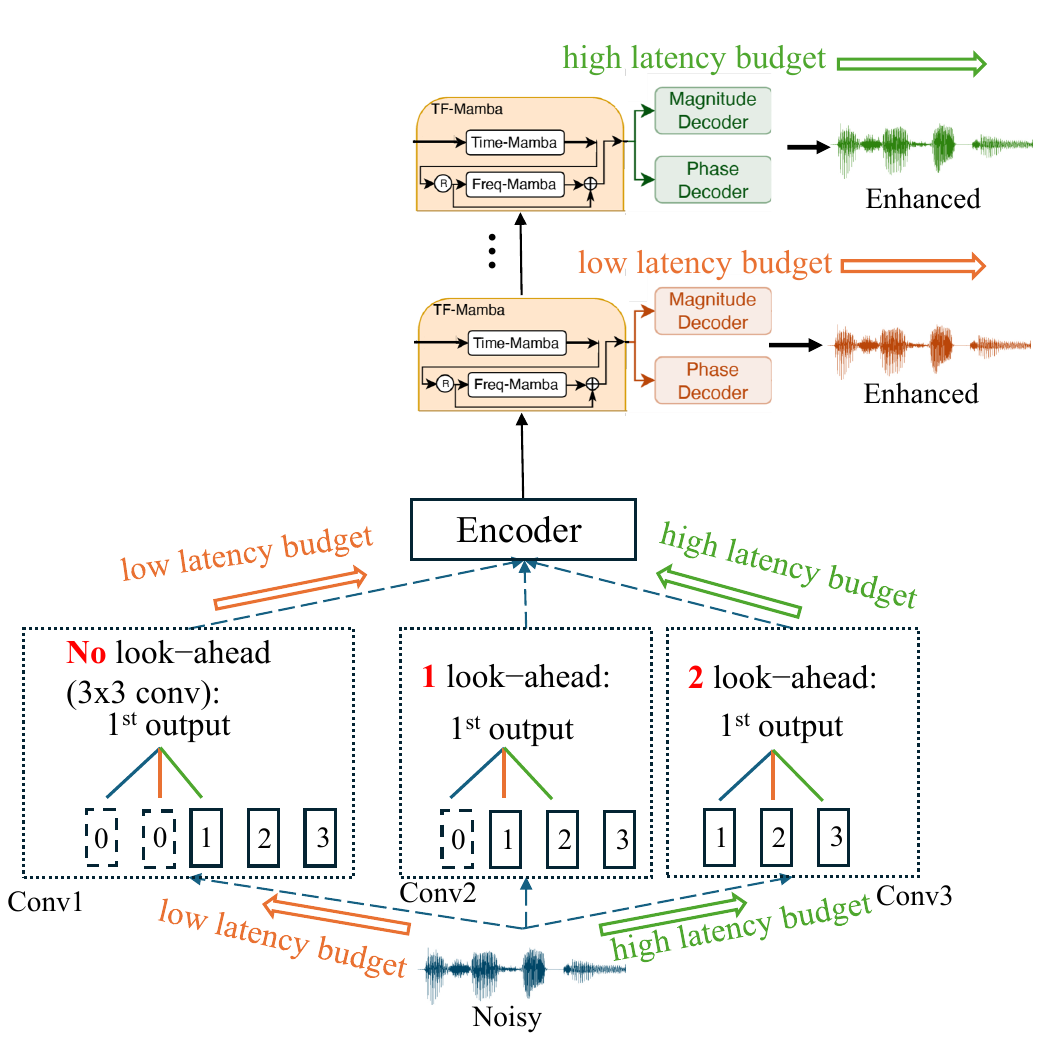}
  \caption{Our proposed one-for-all model enables adjustable algorithmic latency through configurable look-ahead frames and computational latency via early exit. For example, under a low-latency budget, inference can follow the orange arrows.}
  \label{fig:proposed_model}
\end{figure}

To address this issue, inspired by the mixture-of-experts (MoE) paradigm, we employ parallel convolutional layers, each corresponding to a specific number of look-ahead frames (i.e., a particular padding configuration). During training, a convolutional layer is randomly sampled to construct the computational graph. Unlike conventional MoE models, our framework does not require a learned routing mechanism, as the expert selection is explicitly determined by the user based on the latency budget, as illustrated in Fig.~\ref{fig:proposed_model}.

\subsection{Two-Stage Training for Early-Exit Optimization}
\label{Two-Stage}

Although the early-exit mechanism enables flexible inference at different network depths, each intermediate layer represents a compromise, as it must accommodate the requirements of subsequent layers. Consequently, its performance lags behind that of a model optimized for a fixed output depth. One possible remedy is to assign separate decoders to different intermediate layers. However, in the initial experiments on early-exit training, we found that \textbf{sharing a single decoder across all intermediate layers} is more effective, as it enforces a consistent representation space. In contrast, using independent decoders for each layer hinders model learning, as shown in the blue line in Figure \ref{fig:Learnining_curves}.

To address this issue and allow each output layer to learn layer-specific parameters, we adopt a two-stage training framework:  

\begin{enumerate}
\item \textbf{Shared Decoder Stage:} We first train the model with a shared decoder. In each training step, the exit layers are randomly sampled.  

\item \textbf{Multiple Decoder Stage:} After convergence in the first stage, we instantiate independent decoders for each output layer, \textbf{initializing their weights from the shared decoder}. During this stage, the modules preceding the decoders (i.e., the encoder and sequence modeling module) are fine-tuned with a smaller learning rate. 
\end{enumerate}

\textbf{This setting keeps the intermediate layers within a similar representation space while allowing sufficient flexibility to optimize their own outputs.}

\section{Experiments}
\label{exp}

\begin{table*}[t]
    \centering
    \renewcommand{\arraystretch}{0.94}
    \caption{Non-blind test set results of the URGENT 2025 Challenge. Algo. and Comp. denote algorithmic and computational latency, respectively. Algorithmic latency is computed as $40\,\mathrm{ms} + (\#\,\text{look\mbox{-}ahead} \times 20\,\mathrm{ms})$. Computational latency is evaluated on an NVIDIA A100 GPU with 16 kHz input speech. 
    }
    \resizebox{0.99\textwidth}{!}{%
    \begin{tabular}{lcccccccccc}
    \toprule
    \textbf{Method} 
    & \multicolumn{3}{c}{\textbf{Non-intrusive}} 
    & \multicolumn{2}{c}{\textbf{Intrusive}} 
    & \multicolumn{2}{c}{\textbf{Task-ind.}} 
    & \textbf{Task-dep.} 
    & \multicolumn{2}{c}{\textbf{Latency (ms)}} \\
    \cmidrule(lr){2-4}
    \cmidrule(lr){5-6}
    \cmidrule(lr){7-8}
    \cmidrule(lr){9-9}
    \cmidrule(lr){10-11}

    & \textbf{DNSMOS} 
    & \textbf{NISQA} 
    & \textbf{UTMOS}
    & \textbf{PESQ} 
    & \textbf{ESTOI}
    & \textbf{SBERT} 
    & \textbf{LPS}
    & \textbf{CAcc}
    & \textbf{Algo.} 
    & \textbf{Comp.} \\
    \midrule
        
        Noisy & 1.84 & 1.69 &	1.56 &	1.34  &	0.50 
        &	0.74 &	0.61 	
        &	81.29 & - & -\\
        Baseline
        & 2.94 &	2.89 &	2.11 &	- &	- 
        &	- &	- 
        &	84.96 &non-causal &non-causal\\



        \bottomrule
        



        & & & & & \multicolumn{1}{c}{\textbf{\makebox[0pt]{Exit layer=4, Look-ahead=0 (2.0M parameters, MACs (G/s)=19.15)}}} & & & & & \\
        \hdashline
        Specialized (upper bound) & 3.05 & 3.65 & 2.26 & 2.06 & 0.68 
        & 0.82 & 0.76  
        & 82.75 & 40 & 10.09 \\ 


        Early-exit & 3.03 & 3.57 & 2.19 & 2.02 & 0.67 
        & 0.82 & 0.75  
        & 81.13 & 40 & 10.09 \\ 
        

        \textbf{+Parallel conv. (MoE)} &
        2.96 & 3.52 & 2.16 & 2.00 & 0.67 
        & 0.82 & 0.75  
        & 81.69 & 40 & 10.09 \\

        \textbf{ +Multiple dec. stage} &
        2.98 & 3.41 & 2.19 & 2.02 & 0.67 
        & 0.82 & 0.75  
        & 81.86 & 40 & 10.09
        \\
        \bottomrule

        & & & & & \multicolumn{1}{c}{\textbf{\makebox[0pt]{Exit layer=8, Look-ahead=0 (2.9M parameters, MACs (G/s)=25.41)}}} & & & & & \\
        \hdashline
        
        Specialized (upper bound)& 3.10 & 3.77 & 2.36 & 2.19 & 0.71 
        & 0.84 & 0.78  
        & 83.71 & 40 & 18.31  \\

        
        Early-exit & 3.08 & 3.77 & 2.32 & 2.13 & 0.70 
        & 0.83 & 0.77  
        & 81.84  & 40 & 18.31\\ 
        

        \textbf{+Parallel conv. (MoE)} & 3.04 & 3.73 & 2.28 & 2.13 & 0.70 
        & 0.83 & 0.77  
        & 82.72 & 40 & 18.31 \\

        \textbf{ +Multiple dec. stage} & 3.07 & 3.62 & 2.31 & 2.15 & 0.70 
        & 0.84 & 0.77  
        & 83.10 & 40 & 18.31 \\
        
        \bottomrule

        & & & & & \multicolumn{1}{c}{\textbf{\makebox[0pt]{Exit layer=8, Look-ahead=1 (2.9M parameters, MACs (G/s)=25.41)}}} & & & & & \\
        \hdashline
        
        Specialized (upper bound)& 3.15 & 3.90 & 2.42 & 2.27 & 0.72 
        & 0.85 & 0.79  
        & 84.93 & 60 & 18.31 \\

        

        Early-exit  & N/A & N/A & N/A & N/A & N/A 
        & N/A & N/A  
        & N/A  & N/A & N/A\\
        \textbf{+Parallel conv. (MoE)} & 3.10 & 3.82 & 2.34 & 2.21 & 0.71 
        & 0.84 & 0.79  
        & 84.06 & 60 & 18.31 \\

        \textbf{ +Multiple dec. stage} & 3.13 & 3.74 & 2.37 & 2.24 & 0.72 
        & 0.84 & 0.79  
        & 84.62 & 60 & 18.31 \\
        

        \bottomrule
        & & & & & \multicolumn{1}{c}{\textbf{\makebox[0pt]{Exit layer=12, Look-ahead=0 (3.7M parameters, MACs (G/s)=31.67))}}} & & & & & \\
        \hdashline
        
        Specialized (upper bound)& 3.10 & 3.76 & 2.37 & 2.21 & 0.71 
        & 0.84 & 0.78  
        & 84.24 & 40 & 25.05 \\

        
        Early-exit  & 3.10 & 3.81 & 2.34 & 2.14 & 0.70 
        & 0.84 & 0.77  
        & 82.62  & 40 & 25.05\\
        
        
        \textbf{+Parallel conv. (MoE)} & 3.07 & 3.78 & 2.31 & 2.15 & 0.70 
        & 0.83 & 0.77 
        & 82.93 & 40 & 25.05 \\

        \textbf{ +Multiple dec. stage} & 3.11 & 3.70 & 2.34 & 2.17 & 0.70 
        & 0.84 & 0.78  
        & 83.25 & 40 & 25.05 \\
        
        \bottomrule
        
    \end{tabular}
    }
    \label{tab:track1_results}
\end{table*}

\subsection{Dataset}
Following the setup of the URGENT 2025 Challenge~\cite{saijo2025interspeech}, the training dataset consists of multi-condition speech recordings in five languages (English, German, French, Spanish, and Chinese), covering a wide range of sampling rates (8, 16, 22.05, 24, 32, 44.1, and 48 kHz). In addition to clean speech, the dataset includes noise samples and room impulse responses (RIRs). Seven types of degradations are considered: additive noise, reverberation, clipping, bandwidth limitation, codec artifacts, packet loss, and wind noise. The validation set is simulated according to the organizers’ guidelines using the validation splits of the underlying corpora~\cite{saijo2025interspeech}.
The final model checkpoint is selected based on the UTMOS~\cite{saeki2022utmos} score on the validation set. For evaluation, we use the non-blind URGENT 2025 test set, which contains 1,000 utterances.

\subsection{Model Architecture}

Our model architecture largely follows USEMamba~\cite{chao2025universal, chao2024investigation} and RE-USE \cite{fu2025you}, as Mamba~\cite{gu2024mamba} supports RNN-like inference and hence is well-suited for real-time deployment. Considering computational latency constraints, we set the maximum number of Mamba layers to 12 (total 3.7M parameters). To ensure causality or allow only a limited number of look-ahead frames, we introduce the following modifications: (1) replacing standard convolutions with causal convolutions (we explicitly control the number of look-ahead frames by using different amounts of left padding in the first convolutional layer); (2) substituting the bidirectional temporal Mamba with a unidirectional variant; and (3) replacing instance normalization 
with layer normalization 
applied only along the channel dimension.

Unlike RE-USE \cite{fu2025you}, which employs an additional generative model to refine the regression output and achieve a favorable fidelity–quality trade-off, we reduce computational latency by approximating the process of ‘optimally transporting the posterior mean (MMSE estimate) toward the true data distribution’ through a two-stage training strategy: regression loss pre-training followed by adversarial loss fine-tuning guided by a set of discriminators.

To enable a single model to operate across different sampling rates, we adopt sampling frequency-independent (SFI) STFT~\cite{zhang2023toward}, which adjusts the FFT window and hop size according to the input sampling rate while maintaining a fixed time duration. Specifically, we use a 40 ms window and a 20 ms hop for all sampling rates, ensuring an integer number of frequency bins. Therefore, the resulting \textbf{algorithmic latency is 40 ms plus the number of look-ahead frames multiplied by 20 ms}. During training, we randomly sample the exit layer from 3 to 12 and the number of look-ahead frames from 0 to 2. 

We use AdamW with a learning rate of 0.0002 for model training, and set the learning rate to one-tenth of this value for the modules preceding the decoders during the Multiple Decoder Stage.

\subsection{Evaluation Metrics}

To jointly assess perceptual quality and signal fidelity, we employ a diverse set of evaluation metrics. Reference-based metrics include PESQ for perceptual quality~\cite{Rix2001}, ESTOI for intelligibility~\cite{jensen2016stoi}. 
Downstream performance is evaluated using task-independent metrics—SpeechBERTScore (SBERT)~\cite{saeki2024speechbertscore} and Levenshtein Phoneme Similarity (LPS)~\cite{pirklbauer2023evaluation}—as well as task-dependent metrics, 
character accuracy (CAcc) of an ASR ~\cite{peng2024owsm}. Finally, non-intrusive perceptual quality is measured using DNSMOS~\cite{reddy2022dnsmos}, NISQA~\cite{mittag2021nisqa}, and UTMOS~\cite{saeki2022utmos}. Note that, following the setup in RE-USE \cite{fu2025you}, we compute the scores using anechoic clean speech as the reference, rather than early-reflected speech as adopted by the Challenge organizers.

For latency calculation, algorithmic latency is computed as $40\,\mathrm{ms} + (\#\,\text{look\mbox{-}ahead} \times 20\,\mathrm{ms})$, while computational latency is evaluated on an NVIDIA A100 GPU using 16 kHz input speech. We also evaluated the latency on NVIDIA 3090 and 4090 GPUs and observed results within a similar range. Following \cite{welker2025real}, computational latency is measured under online processing (i.e., per-frame computation). However, unlike \cite{welker2025real}, we do not apply \textit{torch.compile} or CUDA graphs to further optimize latency measurement.

\subsection{Results on the non-Blind URGENT 2025 Test Set}

Owing to the flexibility of our proposed one-for-all framework in controlling both algorithmic and computational latency, the model supports 30 different latency configurations, corresponding to 10 exit layers (from 3 to 12) and 3 look-ahead settings (from 0 to 2). In Table~\ref{tab:track1_results}, we present representative results of our method and compare them with noisy speech, the non-causal baseline TF-GridNet~\cite{wang2023tf} from the URGENT 2025 Challenge (which uses early-reflected speech as the learning target), as well as specialized models and early-exit. The specialized model can be regarded as a performance upper bound, but does not provide any latency flexibility. Note that for the 12-layer model, the computational latency is 25 ms, which exceeds the hop size of 20 ms, resulting in an RTF of 1.25. This means it cannot satisfy real-time processing requirements (see Section \ref{Real-Time requirements}) on an NVIDIA A100 GPU, and \textbf{a faster GPU would be needed to achieve real-time performance.}

As shown in Table~\ref{tab:track1_results} and discussed in Section~\ref{Two-Stage}, while early-exit enables flexible control of computational latency, its enhancement performance typically falls short of the specialized model. Our proposed method (\textbf{Parallel conv. (MoE)}), which introduces parallel convolutional layers to control the number of look-ahead frames, achieves performance comparable to early-exit while additionally providing flexibility in controlling algorithmic latency. The proposed two-stage training framework (\textbf{Multiple dec. stage}), incorporating the multiple-decoder stage, consistently improves most evaluation metrics—most notably CAcc (with the exception of NISQA)—and effectively narrows the performance gap with the specialized model.
\textbf{Note that the results for the conventional early-exit model with (Exit Layer = 8, Look-ahead = 1) are marked as N/A, since the conventional early-exit method does not support adjusting the algorithmic latency.}

\begin{figure}[h!]
  \centering

  \begin{subfigure}{\linewidth}
    \centering
    \includegraphics[width=1.0\linewidth]{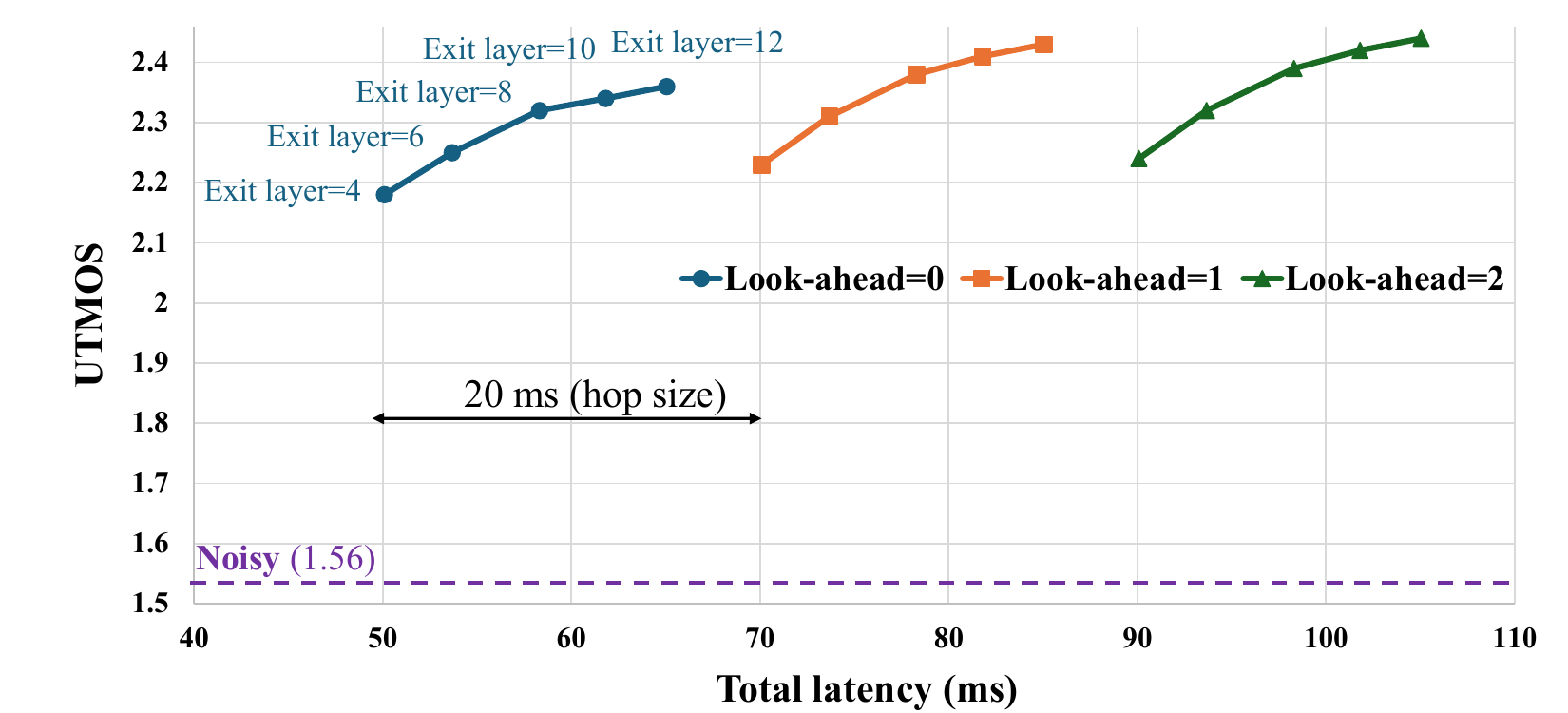}
    \caption{UTMOS vs. Total latency}
    
\vspace{0.5em}

  \begin{subfigure}{\linewidth}
    \centering
    \includegraphics[width=1.0\linewidth]{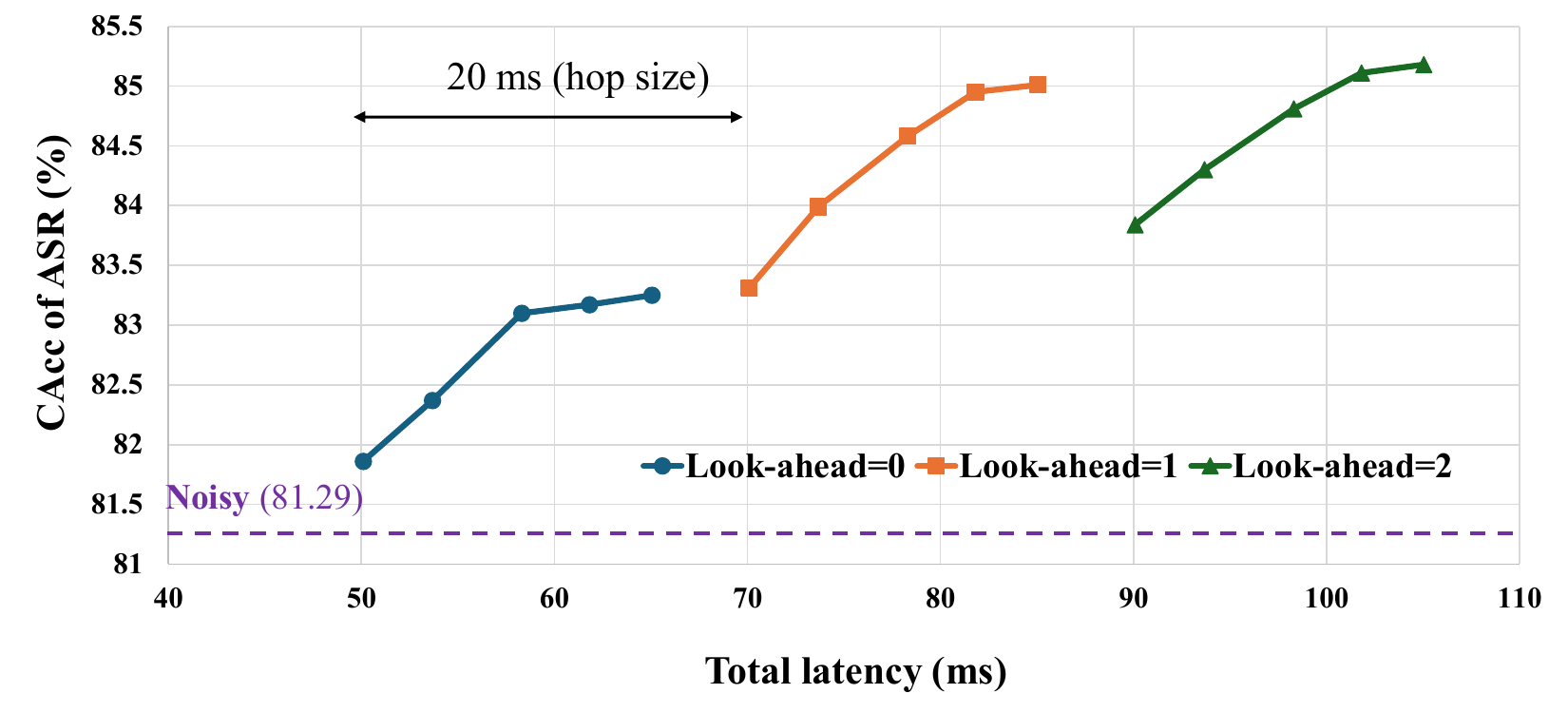}
    \caption{CAcc vs. Total latency}
  \end{subfigure}

  \end{subfigure}
  
  \caption{Relationship between performance metrics and total latency. Our one-for-all model supports 30 distinct latency configurations (results for exit layers 3, 5, 7, 9, and 11 are omitted for brevity).}
  \label{fig:metrics_latency}
\end{figure}

\begin{table*}[t]
    \centering
    \caption{Real-time speech enhancement results on the VoiceBank-DEMAND benchmark. To evaluate generalization across datasets, none of the models (except DEMUCS) are trained on the VoiceBank-DEMAND training set.}
    \renewcommand{\arraystretch}{1.1} 
    \begin{tabular}{lccccc}
        \toprule
        \textbf{Method} & \textbf{PESQ} & \textbf{ESTOI} & \textbf{SI-SDR} &  $\mathbf{\ell_{\text{Algorithm}}}$(ms) & Params
        \\
        \midrule
        Noisy & 1.97 & 0.79 & \phantom{0}8.4 & - & - \\
        
        \midrule
        Diffusion Buffer \cite{lay2025diffusion} & 2.45 & 0.84 & 14.5 &  176 & 22.2M\\
        
        DEMUCS \cite{defossez2020real} & 2.60 & 0.85 & 15.1  & 41 & 33.5M\\
        DeepFilterNet3 \cite{schroter2023deepfilternet} & 2.71 & 0.84 & 17.3 &  40 & 2.14M\\
        
        
        Stream.FM \cite{welker2025real} &  2.72 & 0.85 & 13.4 & 32 & 52.5M
        \\
        \midrule
        \textbf{Proposed (exit layer=8, look-ahead=0)} & 2.76 & 0.86 & 18.6  & 40 & 2.9M
        \\
        \textbf{Proposed (exit layer=8, look-ahead=1)} & \textbf{2.82} & \textbf{0.86} & \textbf{18.8}  & 60 & 2.9M
        \\
        \bottomrule
    \end{tabular}
    \label{tab:VoiceBank}
\end{table*}

In Fig.~\ref{fig:metrics_latency}, we present performance evaluation over a finer grid of total latency. Note that our one-for-all model supports 30 distinct latency configurations; results for exit layers 3, 5, 7, 9, and 11 are omitted for brevity. From the figure, we observe that for UTMOS, performance gains from additional look-ahead frames are less pronounced than those achieved by increasing the model depth. On the other hand, for ASR accuracy, introducing one look-ahead frame substantially improves performance, while adding a second look-ahead frame yields only marginal additional gains.

\subsection{Comparison with Other Real-Time Speech Enhancement Models}

\textbf{To the best of our knowledge, no prior open-source work has proposed a real-time universal speech enhancement model capable of handling complex degradations and inputs with varying sampling rates, as required in the URGENT Challenge setting}. Therefore, to enable comparison with existing real-time speech enhancement models, we evaluate our approach on the widely used VoiceBank-DEMAND benchmark \cite{botinhao2016investigating}, as shown in Table~\ref{tab:VoiceBank}.

For the baselines and to evaluate generalization across datasets, we use DEMUCS \cite{defossez2020real}, DeepFilterNet3 \cite{schroter2023deepfilternet}, Diffusion Buffer \cite{lay2025diffusion}, and Stream.FM \cite{welker2025real}, with most results directly taken from \cite{welker2025real}. Note that, \textbf{consistent with our setting, there is a mismatch between the training and testing data, as none of the compared models (except DEMUCS) are trained on the VoiceBank-DEMAND training set}. For our proposed model, we first set the exit layer to 8 and the look-ahead to 0 to achieve a comparable algorithmic latency to DEMUCS, DeepFilterNet3, and Stream.FM. Compared with these models, our method achieves the highest PESQ, ESTOI, and SI-SDR \cite{le2019sdr}. We also observe that the enhanced speech produced by our model often sounds cleaner than the corresponding clean ground truth. 
In our model, setting the look-ahead to 1 increases the algorithmic latency but consistently improves all evaluation metrics, demonstrating the flexibility of our framework.

\subsection{Practical Deployment}

Users can download the model and evaluate the total latency across different early-exit layers and look-ahead configurations on their own hardware, according to their specific latency budgets. Once the most suitable setting is identified, they can retain only the layers up to the selected exit point and the convolutional branch corresponding to the chosen look-ahead frames. In this way, the resulting model has the same size as a specialized model, without any additional footprint.

Note that computational latency must satisfy the constraints in both Equations~\eqref{eq1} and~\eqref{eq2}, whereas algorithmic latency is governed only by Equation~\eqref{eq1}. Therefore, if users have limited computational resources, they can simply consider increasing the number of look-ahead frames.

\section{Future work}

In this paper, we focus on training a flexible one-for-all model that can be readily deployed under diverse conditions. Techniques such as pruning and quantization offer promising directions for accelerating inference and can be naturally combined with our framework. Another avenue for future work is to further reduce the performance gap between shallow and deep outputs. In particular, knowledge distillation strategies, inspired by recent large-to-small language model compression, 
 may be explored to enhance the performance of shallower exits.

\section{Conclusion}

We propose a one-for-all, real-time universal speech enhancement framework that explicitly controls both algorithmic and computational latency within a single model. By introducing parallel convolutional layers, we enable flexible adjustment of look-ahead frames for algorithmic latency control, while the early-exit mechanism allows dynamic control of computational latency through variable network depth. To mitigate the performance gap between flexible and specialized models, we further developed a two-stage training strategy with a shared-to-multiple decoder transition, which effectively stabilizes learning and improves intermediate-layer performance. Experimental results on the URGENT 2025 Challenge dataset demonstrate that the proposed framework supports 30 distinct latency configurations while maintaining performance close to specialized models. These results show that a single model can adapt to diverse real-time applications without retraining separate models.

\section{Generative AI Use Disclosure}
Generative AI was used only for editing and polishing this manuscript.

\newpage
\bibliographystyle{IEEEtran}
\bibliography{Interspeech}

\end{document}